\documentclass[journal]{IEEEtran}

\ifCLASSINFOpdf
\else
   \usepackage[dvips]{graphicx}
\fi

\usepackage{url}
\usepackage{textcomp}
\usepackage{stfloats}
\usepackage{hyperref} 
\usepackage{subfigure}
\usepackage{verbatim}
\usepackage{makecell}
\usepackage{multirow}

\usepackage{graphicx}
\usepackage{cite}
\usepackage{amssymb}
\usepackage{color}
\usepackage{xcolor}
\usepackage{amsmath,amsfonts}
\usepackage{algorithmic}
\usepackage{algorithm}
\usepackage{array}
\usepackage{booktabs}
\begin{document}

\title{StreamVoice+: Evolving into End-to-end Streaming Zero-shot Voice Conversion}



\author{Zhichao~Wang,
    Yuanzhe~Chen,
    Xinsheng~Wang,
    Lei~Xie,~\IEEEmembership{Senior Member, 
IEEE},
    and Yuping Wang

\thanks{Zhichao Wang, Xinsheng Wang, and Lei Xie are with the ASLP Lab, School of Computer Science, Northwestern Polytechnical University, Xi’an 710129, China (email: zcwang\_aslp@mail.nwpu.edu.cn; w.xinshawn@gmail.com; lxie@nwpu.edu.cn)}
\thanks{Yuanzhe Chen and Yuping Wang are with the ByteDance, China (email: chenyuanzhe@bytedance.com; wangyuping@bytedance.com)}
    }

\markboth{Journal of \LaTeX\ Class Files, Vol. 14, No. 8, August 2015}
{Shell \MakeLowercase{\textit{et al.}}: Bare Demo of IEEEtran.cls for IEEE Journals}
\maketitle

\begin{abstract}

StreamVoice has recently pushed the boundaries of zero-shot voice conversion (VC) in the streaming domain. It uses a streamable language model (LM) with a context-aware approach to convert semantic features from automatic speech recognition (ASR) into acoustic features with the desired speaker timbre. Despite its innovations, StreamVoice faces challenges due to its dependency on a streaming ASR within a cascaded framework, which complicates system deployment and optimization, affects VC system's design and performance based on the choice of ASR, and struggles with conversion stability when faced with low-quality semantic inputs. To overcome these limitations, we introduce StreamVoice+, an enhanced LM-based end-to-end streaming framework that operates independently of streaming ASR. StreamVoice+ integrates a semantic encoder and a connector with the original StreamVoice framework, now trained using a non-streaming ASR.  This model undergoes a two-stage training process: initially, the StreamVoice backbone is pre-trained for voice conversion and the semantic encoder for robust semantic extraction. Subsequently, the system is fine-tuned end-to-end, incorporating a LoRA matrix to activate comprehensive streaming functionality. Furthermore, StreamVoice+ mainly introduces two strategic enhancements to boost conversion quality: a residual compensation mechanism in the connector to ensure effective semantic transmission and a self-refinement strategy that leverages pseudo-parallel speech pairs generated by the conversion backbone to improve speech decoupling. Experiments demonstrate that StreamVoice+ not only achieves higher naturalness and speaker similarity in voice conversion than its predecessor but also provides versatile support for both streaming and non-streaming conversion scenarios.

\end{abstract}

\IEEEpeerreviewmaketitle

\vspace{-10pt}
\section{Introduction}

\IEEEPARstart{V}{oice} conversion (VC) aims to convert a speaker's voice to that of a target speaker without altering the linguistic content. This technique is applied in various real-world scenarios, e.g., movie dubbing, privacy protection, pronunciation correction, etc. Meanwhile, zero-shot VC, which enables conversion to any target speaker using only one utterance from that speaker, has drawn much attention~\cite{autovcqian2019autovc,LM-VC}.  However, the increasing demand for streaming capabilities in real-time applications, like live broadcasting and online meetings, poses a new challenge to zero-shot VC, which mainly focuses on offline processing. This letter focuses on \textit{streaming zero-shot VC}, which performs real-time conversion given any speakers.

To achieve streaming capability, causal processing and streamable structures are essential. However, the typical absence of future information in streaming models may inevitably degrade performance. Common approaches to mitigate this degradation include enhancing semantic information~\cite{YZ,chen2022streaming,dualvc3} and distilling knowledge~\cite{dualvc,dualvc2,dualvc3,fasts2s,investgateStreaming} from a non-streaming model through parameter sharing or guided training. Nonetheless, previous streaming VC methods mainly focus on pre-defined speakers. To enable conversion for any speaker, the fundamental strategy involves the disentanglement and recombination of speech components, e.g. semantic content and speaker timbre, employing either pre-training techniques~\cite{PPGSun2016PhoneticPF,autovcqian2019autovc,nansy,mediumvc,wang2023multilevel} or jointly trained encoders~\cite{INchou2019oneshot,VQMIVC,contrastive,avqvc}. Recent advancements~\cite{liveosvc,ALO-VC,streamvoice} integrate streaming capability with zero-shot VC, either by adapting non-streaming models to be streamable or by using streaming pre-trained models like ASR and speaker verification (SV) models.



Inspired by the success of LM-based VC models~\cite{makeavoice,uniaudio,LM-VC} in offline conversion, LM-based StreamVoice~\cite{streamvoice} represents a significant advancement in streaming zero-shot VC. Building on a recognition-synthesis framework~\cite{PPGSun2016PhoneticPF}, where speech is represented as semantic and acoustic features extracted via a streaming ASR and an audio codec respectively, StreamVoice transforms source semantic information into acoustic features with the target speaker's timbre. To enhance historical context learning and anticipate missing future information, semantic masking and teacher-guided context foresight are employed in StreamVoice. Despite its good zero-shot performance, StreamVoice shows a strong dependency on cascaded streaming pipelines leading to several disadvantages. 1) \textit{Complexity}: The integration of multiple models with various structures complicates optimization and deployment. 2) \textit{Flexibility}: The choice of streaming ASR affects VC design and performance, limiting implementation and further extension flexibility. 3) \textit{Stability}: Low-quality semantic information from streaming ASR, which may include unexpected speaker timbre and noise, leads to unstable conversion for diverse inputs.

To overcome these issues in StreamVoice, we propose StreamVoice+, a concise LM-based streaming framework for ASR-free end-to-end zero-shot VC. Inspired by the capability for modality or embedding alignment via backbone pre-training~\&~task-orient fine-tuning paradigm~\cite{PEFT,tang2024salmonn,ma2024embarrassingly}, 
the core concept behind StreamVoice+ involves leveraging a high-performance StreamVoice, originally trained on a non-streaming ASR, as the foundational model. We extend the capabilities of this backbone by attaching a semantic encoder and a connector via end-to-end fine-tuning. This integration facilitates an ASR-free, end-to-end streaming conversion. To be specific, StreamVoice+ employs a two-stage training process: first, pre-training the StreamVoice backbone for conversion and a semantic encoder for semantic extraction using high-quality semantic information from a non-streaming ASR, and then fine-tuning the entire model with additional LoRA adapters to unlock end-to-end conversion capability. To enhance decoupling and conversion quality, we mainly introduce two strategies: 1) \textbf{r}esidual compensation with a \textbf{b}ottleneck in the connector, referred to as the R-B connector, which is designed to ensure the transmission of semantic content while minimizing the influence of the source speaker's timbre; and 2) a self-refinement strategy that uses pseudo-parallel speech pairs generated by the backbone to aid in decoupling training. Experimental results show that StreamVoice+ achieves end-to-end streaming conversion with superior performance compared to StreamVoice. The total pipeline latency is 112 ms, making it 1.7x faster than real-time on a single A100 GPU without engineering optimizations. We also show that StreamVoice+ can easily be extended to support non-streaming and streaming conversion with simple modifications. Converted samples can be found in \href{https://kerwinchao.github.io/StreamVoice_Plus/}{\url{https://kerwinchao.github.io/StreamVoice_Plus/}}.


\begin{figure*}[ht]
\centering
\begin{minipage}{0.43\linewidth}
\centering
    \subfigure[StreamVoice+]{
      \includegraphics[width=1\columnwidth]{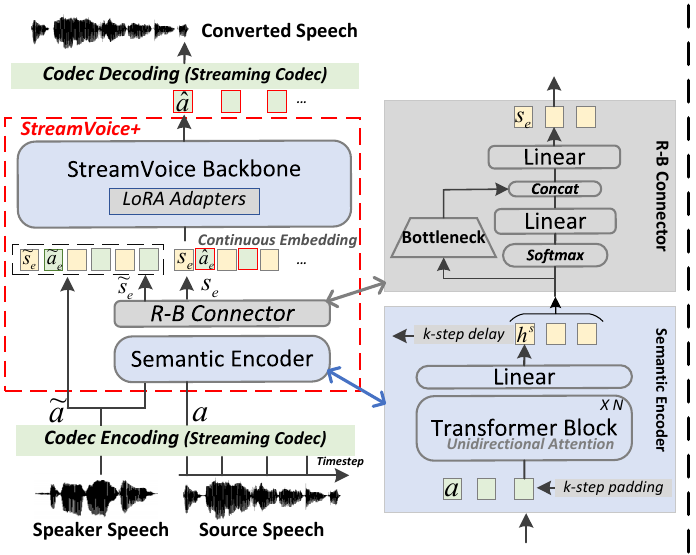}}
\end{minipage}
\hspace{0.5cm}
\begin{minipage}{0.22\linewidth}
\centering
    \subfigure[Training Stage: Pre-training]{
      \includegraphics[width=1\columnwidth]{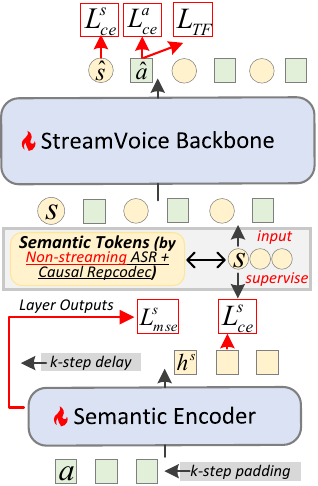}}
\end{minipage}
\hspace{0.5cm}
\begin{minipage}{0.24\linewidth}
\centering
    \subfigure[Training Stage: Fine-tuning]{
      \includegraphics[width=1\columnwidth]{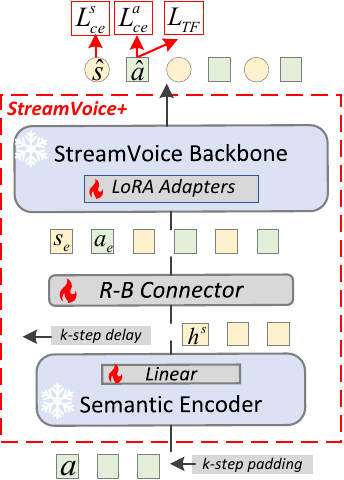}}
\end{minipage}
\vspace{-5pt}
\caption{The framework of (a) StreamVoice+, which employs two-stage training procedura: (b) pre-training and (c) fine-tuning, to achieve end-to-end conversion.}
\label{fig:zsl_framework}
\vspace{-13pt}
\end{figure*}

\vspace{-10pt}
\section{Proposed Approach}

\subsection{Architecture of StreamVoice+}
\label{sec:structure}
As shown in Fig.~\ref{fig:zsl_framework}a, StreamVoice+ consists of a semantic encoder, a connector, and a StreamVoice backbone with built-in LoRA adapters~\cite{lora}.
In this framework, speech is represented as acoustic feature $\mathbf{a} \in \mathbb{R}^{T\times L}$ by a speech codec~\cite{audiodec}, where $T$ denotes the sequence length and $L$ represents the number of quantizers in the codec. 
With the acoustic feature $\mathbf{\tilde{a}}$ from the target speaker, StreamVoice+ casually converts source acoustic feature $\mathbf{a}$ to the output $\hat{\mathbf{a}}$. To be specific, $\mathbf{\tilde{a}}$ and $\mathbf{a}$ are first processed by the semantic encoder and connector to extract the continuous semantic information $\mathbf{\tilde{s}_e}$ and $\mathbf{s_e}$. Using the speaker prompt $\{\mathbf{\tilde{s}},\mathbf{\tilde{a}}\}$, StreamVoice backbone transforms the source semantic information $\mathbf{s_e}$ into final output $\hat{\mathbf{a}}$.

\subsubsection{Semantic Encoder}
As shown in Fig.~\ref{fig:zsl_framework}a, the semantic encoder extracts the semantic information from the acoustic input, resulting in the hidden semantic output $\mathbf{h}^s$. To meet the streaming requirement, the encoder is achieved by $N$-layer Transformer blocks~\cite{transformer} with unidirectional attention and a linear projection. Additionally, a behavior of k-step output delay is introduced in the encoder to generate $h_t^s$ when getting future $k$-step acoustic input $\mathbf{a}_{t:t+k}$, achieving a better trade-off between latency and performance of semantic extraction.

\subsubsection{StreamVoice Backbone with LoRA}
Inspired by the recent LM advance~\cite{tang2024salmonn,ma2024embarrassingly,vediollama2}, StreamVoice+ adapts the pre-trained StreamVoice~\cite{streamvoice} as its conversion backbone and extends the capability of StreamVoice with LoRA adapters for end-to-end conversion, in which LoRA is only plugged into the key, query, and value projections in the self-attention mechanism. Following the original version~\cite{streamvoice}, StreamVoice backbone integrates a full causal LM that generates acoustic codecs in collaboration with an acoustic predictor. By alternating the input of semantic and acoustic features at each time step, StreamVoice ensures streaming behavior.

\subsubsection{R(esidual)-B(ottleneck) Connector}
The connector transforms the output from the semantic encoder into embeddings that are compatible with the backbone model, as shown in Fig.~\ref{fig:zsl_framework}a. In other words, the linear-based connector ensures that the semantic output $\mathbf{h}^s$ from the semantic encoder closely approximates the original semantic input $\mathbf{s}$ of StreamVoice in the continuous embedding space. Since $\mathbf{h}^s$ contains not only semantic information but also undesired speaker timbre, a softmax function, which is also used in CE-based optimization of the semantic encoder (See Section~\ref{sec:traing}), is applied in the connector. However, this normalization may compromise the semantic content. To mitigate this effect and enhance semantic quality, the connector is designed to deliver residual information via a bottleneck employing a skip connection way.
This residual information is subsequently combined with the main branch to compensate for the loss in semantic information, resulting in $\mathbf{s_e}$.

\vspace{-12pt}
\subsection{Two-stage Training Procedure}
\label{sec:traing}
In StreamVoice+, the basic idea is to employ a high-performance backbone model, such as non-streaming StreamVoice, and then augment its functionality by attaching a semantic encoder and a connector via end-to-end fine-tuning. This integration enables ASR-free end-to-end streaming conversion. StreamVoice+ employs a two-stage training procedure, which includes pre-training and end-to-end fine-tuning, as described below.



\subsubsection{Pre-training}
As shown in Fig.~\ref{fig:zsl_framework}b, we separately train StreamVoice for conversion and semantic encoder for semantic extraction, using high-quality semantic information from a non-streaming ASR as input or supervision. 
In this stage, discrete semantic feature $\mathbf{s} \in \mathbb{R}^{T \times 1}$ is extracted from the speech utterance. Here, non-streaming ASR aggregates the continuous semantic information, which is then discretized by modified RepCode~\cite{repcodec} with \textit{causal} convolution. For the StreamVoice backbone, we follow the original configuration~\cite{streamvoice}, which minimizes the cross entropy (CE) loss for codec prediction $\mathcal{L}^{a}_{ce}$ and teacher foresight loss $\mathcal{L}_{TF}$. Additionally, semantic prediction loss $\mathcal{L}^{s}_{ce}$ is also implemented to enhance performance.
For the semantic encoder, we optimize it with CE loss $\mathcal{L}^{s}_{ce}$ for semantic prediction. We also introduce intermediate layer supervision~\cite{intermedium}, which uses the continuous semantic feature to supervise the layer outputs by mean square error (MSE) $\mathcal{L}^{s}_{mse}$, encouraging intermediate layers to learn semantic knowledge. The total loss in pre-training can be defined as
$\mathcal{L}_{Backbone}=\mathcal{L}^{s}_{ce}+\mathcal{L}^{a}_{ce}+\mathcal{L}_{TF}$ and $\mathcal{L}_{Encoder}=\mathcal{L}^{s}_{ce}+\mathcal{L}^{s}_{mse}$.




\subsubsection{Fine-tuning with self-refinement strategy}
Integrating the pre-trained capabilities into StreamVoice+, we fine-tune the whole model with LoRA adapters to unlock the end-to-end conversion ability. As presented in Fig.~\ref{fig:zsl_framework}c, the semantic encoder and backbone are frozen in this stage, while the connector and LoRA are responsible for the end-to-end training. Following the attempt in SpearTTS~\cite{spearTTS}, the last linear layer in the semantic encoder is also optimized for better performance. This fine-tuning only updates \textit{3.9M} parameters of \textit{153M} StreamVoice+. Generally, in VC, 
a single speech utterance is used simultaneously as source and target speech, which may encourage the model to focus solely on reconstruction, thereby neglecting speech decoupling and resulting in poor conversion stability. This issue is exacerbated in end-to-end training.
To mitigate this issue, the intuitive way is to create a parallel speech pair with the same content but different speaker timbre. To this end, we introduce a self-refinement strategy, which employs the pre-trained backbone to perform conversion on the training dataset for creating parallel pairs. In practice, we randomly replace the source or target speech with the synthetic speech and optimize StreamVoice+ in both conversion and reconstruction behavior. In the fine-tuning stage, StreamVoice+ is only optimized by $\mathcal{L}_{Backbone}$.

\vspace{-10pt}
\subsection{Dual-mode Extension: Streaming \& Non-streaming}
\vspace{-2pt}
With the architecture of StreamVoice+, there is a straightforward way to unify streaming and non-streaming conversion into a single framework, which can reduce the cost of practical applications. The core idea involves the integration of task-specific parameters into StreamVoice+.
Specifically, after the pre-training stage, we further freeze the semantic encoder and continue training the semantic encoder for non-streaming scenarios by incorporating \textit{bidirectional} attention and additional LoRA parameters. Then, similar to the fine-tuning stage described in Section~\ref{sec:traing}, another set of linear layer, connector, and LoRA adapters of the backbone are optimized for the non-streaming conversion task. Compared with the original StreamVoice+, only an additional \textit{4.8M} task-specific parameters need to be stored.

\vspace{-5pt}
\section{Experiments}
\label{sec:exp}
\vspace{-3pt}
\subsection{Experimental Setup}
\vspace{-2pt}
\subsubsection{Corpus}
WenetSpeech4TTS Basic~\cite{wenetspeech4tts}, Aishell3~\cite{aishell3} and an internal dataset, in total 8,700 hours of 16kHz recordings, are used to train StreamVoice+, Repcodec~\cite{repcodec}, and Audiodec~\cite{audiodec}. For semantic extraction, we incorporate a non-streaming ASR\footnote{https://github.com/wenet-e2e/wenet/tree/main/examples/wenetspeech/s0\label{model:nsasr-wenetspeech}} trained on WenetSpeech~\cite{wenetspeech}. For testing, 600 testing pairs are selected from DIDISpeech~\cite{didispeech}, EMIME~\cite{emime}, and an internal dataset, each with a source and target speaker utterance. During inference, a 3s speaker prompt is used. The duration of testing utterances ranges between 3s and 7s.

\subsubsection{Details}
Audiodec\footnote{https://github.com/facebookresearch/AudioDec} we used has 4 quantizers with a 1024 codebook, representing a 24kHz waveform in 20ms frame length. The ASR model and RepCodec\footnote{https://github.com/mct10/RepCodec} extract a discrete semantic feature with a 40ms frame length. For StreamVoice+, the StreamVoice backbone uses the original configuration~\cite{streamvoice} containing 6-layer LLaMA~\cite{llama} and a single Transformer-layer acoustic predictor. Semantic masking of StreamVoice is also kept in training. The LoRA adapters in the backbone use 32-rank and $\alpha=1$. The semantic encoder is implemented by a 3-layer LLaMA with 8 heads. The hidden and intermediate sizes are 1024 and 4096. The connector's unit size is 1024 and the bottleneck dimension is 16. The delay is set to 80ms (k=4). Eight V100 GPUs are used to pre-train the backbone and semantic encoder for 500k steps. We use the AdamW optimizer with a learning rate of $5 \times 10^{-4}$. Exponential decay updates the learning rate after each epoch. During fine-tuning, StreamVoice+ is trained for 100k steps with $4 \times 10^{-4}$ learning rate with a decay period of 10k steps.


\subsubsection{Evaluation metrics}
The mean opinion score (MOS) subjectively measures speech naturalness (NMOS) and speaker similarity (SMOS), calculated with 95$\%$ confidence intervals. We randomly select 120 testing pairs for subjective evaluations involving a group of 15 listeners. For objective evaluations, a neural network-based system\footnote{https://github.com/AndreevP/wvmos} is used to measure speech quality (WVMOS). Character error rate (CER) measured by an ASR model~\footref{model:nsasr-wenetspeech} indicates speech intelligibility. 
Pearson correlation coefficient (PCC) of fundamental frequency measures the speaking style reservation after conversion. Speaker cosine similarity (SSIM) is calculated by an SV model~\cite{wespeaker} to determine if the converted speech matches the target speaker.


\vspace{-15pt}
\subsection{Zero-shot Evaluation}
\vspace{-3pt}
We select one LM-based zero-shot VC system, \textit{LM-VC}~\cite{LM-VC}, as the topline system. Additionally, we compare the non-streaming \textit{backbone} model of StreamVoice+ and the streaming predecessor StreamVoice~\cite{streamvoice}. We implement the proposed system \textit{StreamVoice+}, and also involve the non-streaming part of the dual-mode \textit{N-StreamVoice+} in the evaluation. Please note that all comparison systems are trained on the same dataset. Table~\ref{exp:zeroshot} presents both subjective and objective results. As we can see, compared with the non-streaming LM-VC and backbone model, StreamVoice+ can achieve comparable results in terms of subjective NMOS and SMOS. From the aspect of objective results, there is still a performance gap in CER and SSIM between StreamVoice+ and the non-streaming models, and meanwhile, StreamVoice+ demonstrates superior performance in NMOS and PCC, benefiting from end-to-end training. Additionally, the non-streaming model of dual-mode StreamVoice+ even surpasses the topline models in most aspects, indicating the effectiveness of our end-to-end streaming framework. Among the streaming models,  StreamVoice+ surpasses StreamVoice in almost all metrics, although CER is slightly lower. This can be attributed to the robustness of semantic extraction, which can be mitigated by speech augmentation and scaling up the dataset.
Real-time factor (RTF) of StreamVoice+ and codec are 0.58 and 0.004, meeting the real-time requirement. With an 80ms designed delay and a 20ms token length, the overall pipeline latency of StreamVoice+ is $111.6ms = 80 + 20 + 20 \times (0.58 + 0.004)$ on an A100 GPU. Compared with StreamVoice ($124.3ms$), StreamVoice+ is built on an end-to-end framework and has a concise streaming pipeline without dependency on streaming ASR. These results demonstrate the powerful capability of StreamVoice+ in streaming zero-shot VC.



\begin{table}[htp]
\vspace{-10pt}
\caption{Zero-shot performance}
\vspace{-5pt}
\label{exp:zeroshot}
\footnotesize
\setlength{\tabcolsep}{0.8mm}
\renewcommand\arraystretch{1.22}
\begin{tabular}{lccccccc}
\hline
\multicolumn{1}{l}{\multirow{2}{*}{Method}} & \multicolumn{4}{c}{Conversion Quality} & \multicolumn{2}{c}{Speaker Similarity}  \\ \cline{2-7}
  &  \scriptsize{NMOS} $\uparrow$  & \scriptsize{WVMOS} $\uparrow$  & \scriptsize{CER}  $\downarrow$ & \scriptsize{PCC}  $\uparrow$ & \scriptsize{SMOS} $\uparrow$ & \scriptsize{SSIM} $\uparrow$ \\ \hline
 GT~(origin) & -  & 3.61  & 6.29 & -  & - &  0.853   \\ \hline
 \multicolumn{2}{l}{\textit{\textcolor[RGB]{57,57,57}{Non-streaming Topline}}}    &   &   &  &     \\  
 LM-VC & 3.70$\pm$0.08  & 3.58  &  9.50 & 0.532 & 3.73$\pm$0.06   & 0.776   \\
 Backbone & 3.78$\pm$0.05  & 3.65  & \textbf{8.61} & 0.565 & \textbf{3.82$\pm$0.08}  &  0.781  \\
 N-StreamVoice+ & \textbf{3.81$\pm$0.06}  & \textbf{3.75}  & 9.69 & \textbf{0.611} & 3.79$\pm$0.05  & \textbf{0.783}   \\ \hline
 \multicolumn{2}{l}{\textit{\textcolor[RGB]{57,57,57}{Streaming Model}}}    &   &   &  &     \\ 
 StreamVoice & 3.71$\pm$0.08  & 3.63  & \textbf{10.1} & 0.591 & 3.68$\pm$0.07  & 0.758  \\
  StreamVoice+ & \textbf{3.75$\pm$0.06} & \textbf{3.74} & 10.8 &  \textbf{0.632}    & \textbf{3.75$\pm$0.06}  & \textbf{0.776}  \\ \hline
\end{tabular}
\end{table}

\vspace{-20pt}
\subsection{Component Analysis}
To get insight into StreamVoice+, we further analyze the key configurations of model components and training. 
\subsubsection{Semantic Encoder}
We observe that the designed k-step delay and the number of LLaMa layers affect the conversion performance in practice, as shown in Table~\ref{exp:abs}. For the k-step delay, a higher built-in delay enables the encoder to achieve lower pre-training loss and better conversion performance. Conversely, discarding the delay (k=0) leads to rapid performance degradation. This can be attributed to the inherent reception of semantic features from non-streaming ASR, which may capture future semantic information. Additionally, using more layers in the semantic encoder results in better conversion performance but increases the RTF.

\subsubsection{R-B Connector}
Compared with the generally used linear connector, we use the R-B connector with residual compensation for better decoupling ability. When the residual path is removed (dim=0), SSIM shows a slight improvement, but there is a significant drop in CER, indicating the importance of semantic compensation provided by the residual design. Conversely, with an excessively large bottleneck (dim=64), unexpected speaker timbre and noise leads to decreased speaker similarity and speech quality.

\subsubsection{Training Procedure}
In StreamVoice+, the two-stage training procedure is the key to the end-to-end framework. 
During pre-training, the $\mathcal{L}_{Encoder}$ enables the semantic encoder to aggregate semantic information, narrowing the distance with the semantic space of the backbone. When such pre-training is canceled, StreamVoice+ struggles to achieve high-quality conversion. With semantic knowledge learned in pre-training, freezing the encoder during the end-to-end fine-tuning stage hinders the benefits of end-to-end optimization. Also, using a self-refinement strategy can effectively prompt speech decoupling and create a conversion simulation during training. For LoRA adapters, when dropping the LoRA, the frozen backbone causes a semantic mismatch with the semantic encoder, leading to lower CER and PCC. Besides, fully tuning the backbone can cause it to drift significantly from its original parameters, harming conversion ability. These findings indicate that the employment of LoRA is effective in fine-tuning without compromising its performance. This pluggable module is also advantageous for extending dual-mode conversion and facilitating deployment.

\subsubsection{Dataset Size}

 
In addition to training StreamVoice+ on the 8700-hour dataset, we also implement two versions using 1500-hour and 5500-hour subsets from the original dataset. As shown in Table~\ref{exp:zeroshot}, there is a clear trend that with more training data, StreamVoice+ achieves higher performance, particularly in CER and SSIM. For the end-to-end framework, the scale of training data is crucial to its performance and robustness. We believe that StreamVoice+ can mitigate the gaps in speech intelligibility and speaker similarity by using more training data, thereby achieving more powerful conversion capabilities.


\begin{table}[htp]
\footnotesize
\vspace{-10pt}
\caption{Results of ablation studies.}
\vspace{-5pt}
\setlength{\tabcolsep}{1.5mm}
\renewcommand\arraystretch{1}
\centering
\begin{tabular}{lcccc}
\hline
 Method   & WVMOS $\uparrow$  & CER  $\downarrow$ & PCC  $\uparrow$  & SSIM $\uparrow$ \\ \hline
  StreamVoice+  &  3.74 & 10.8 &  0.632    & 0.776 \\ \hline
 \multicolumn{3}{l}{ \textit{\textcolor[RGB]{57,57,57}{Semantic Encoder (4-step delay, 3 layers)}}}           &   \\ 
   $\quad$K-step Delay: 0    & 3.54  & 41.3   & 0.617  & 0.748   \\
$\quad$\textcolor{white}{laddd}$\rightarrow$ 2 ($=40ms$)   & 3.77 & 31.0   & 0.602 & 0.772   \\
$\quad$Layer Num: 2    & 3.67  & 14.9   & 0.628  & 0.764   \\
$\quad$\textcolor{white}{layerdd d}$\rightarrow$ 6   & 3.76 & 10.2   & 0.628 & 0.783   \\\hline
 \multicolumn{3}{l}{ \textit{\textcolor[RGB]{57,57,57}{R-B Connector (16 residual dim) }}}           &   \\ 
$\quad$Residual Dim: 0    & 3.82  & 22.3   & 0.609  & 0.789   \\
$\quad$\textcolor{white}{Residual d}$\rightarrow$ 64   &3.60  & 14.2   & 0.643 & 0.750   \\ \hline
 \multicolumn{4}{l}{ \textit{\textcolor[RGB]{57,57,57}{Pre-training (P) + Fine-tuning (F) Procedure}}}           &   \\ 
 $\quad$\textit{P} \textit{w/o} Encoder   & 3.37 & 9.49  & 0.695 & 0.586  \\
 $\quad$\textit{F} \textit{w/} Frozen Encoder   & 3.81 & 13.5  & 0.616 & 0.778  \\
 $\quad$\textit{F} \textit{w/o} Self Refinement   & 3.62  & 10.7   & 0.645  & 0.743   \\
\multicolumn{3}{l}{$\quad$\textit{F} \textit{w/o} LoRA: } &   \\

$\quad$$\quad$$\rightarrow$ Frozen Backbone   & 3.73  & 26.5   & 0.533  & 0.778   \\
$\quad$$\quad$$\rightarrow$ Trainable Backbone  & 3.33 & 14.4   & 0.650 & 0.709   \\ \hline
 \multicolumn{4}{l}{ \textit{\textcolor[RGB]{57,57,57}{Dataset Size (8700h)}}}           &   \\ 
$\quad$1500h   & 3.63  & 37.5   & 0.570  & 0.719   \\
$\quad$5500h    & 3.73  & 19.3   & 0.634  & 0.759   \\  \hline
\end{tabular}
\label{exp:abs}
\end{table}

\vspace{-15pt}
\section{CONCLUSIONS}
This letter discusses the task of streaming zero-shot VC and proposes an LM-based end-to-end framework StreamVoice+, which aims to solve the problems of complexity, inflexibility, and instability caused by strong dependencies in StreamVoice. StreamVoice+ extends the capabilities of a high-performance non-streaming conversion backbone by attaching a semantic encoder and a connector via two-stage training, enabling ASR-free end-to-end streaming conversion. Experiments show the end-to-end streaming conversion ability of StreamVoice+, which obtains superior naturalness and speaker similarity to its predecessor StreamVoice.

\newpage

\bibliographystyle{IEEEtran}
\bibliography{ref.bib}

\begin{thebibliography}{10}
\providecommand{\url}[1]{#1}
\csname url@samestyle\endcsname
\providecommand{\newblock}{\relax}
\providecommand{\bibinfo}[2]{#2}
\providecommand{\BIBentrySTDinterwordspacing}{\spaceskip=0pt\relax}
\providecommand{\BIBentryALTinterwordstretchfactor}{4}
\providecommand{\BIBentryALTinterwordspacing}{\spaceskip=\fontdimen2\font plus
\BIBentryALTinterwordstretchfactor\fontdimen3\font minus \fontdimen4\font\relax}
\providecommand{\BIBforeignlanguage}[2]{{%
\expandafter\ifx\csname l@#1\endcsname\relax
\typeout{** WARNING: IEEEtran.bst: No hyphenation pattern has been}%
\typeout{** loaded for the language `#1'. Using the pattern for}%
\typeout{** the default language instead.}%
\else
\language=\csname l@#1\endcsname
\fi
#2}}
\providecommand{\BIBdecl}{\relax}
\BIBdecl

\bibitem{autovcqian2019autovc}
K.~Qian, Y.~Zhang, S.~Chang, X.~Yang, and M.~Hasegawa-Johnson, ``Autovc: Zero-shot voice style transfer with only autoencoder loss,'' in \emph{International Conference on Machine Learning (ICML)}, 2019, pp. 5210--5219.

\bibitem{LM-VC}
Z.~Wang, Y.~Chen, L.~Xie, Q.~Tian, and Y.~Wang, ``Lm-vc: Zero-shot voice conversion via speech generation based on language models,'' \emph{IEEE Signal Processing Letters}, pp. 1157--1161, 2023.

\bibitem{YZ}
Y.~Chen, M.~Tu, T.~Li, X.~Li, Q.~Kong, J.~Li, Z.~Wang, Q.~Tian, Y.~Wang, and Y.~Wang, ``Streaming voice conversion via intermediate bottleneck features and non-streaming teacher guidance,'' in \emph{International Conference on Acoustics, Speech and Signal Processing (ICASSP)}, 2023, pp. 1--5.

\bibitem{chen2022streaming}
Z.~Chen, H.~Miao, and P.~Zhang, ``Streaming non-autoregressive model for any-to-many voice conversion,'' \emph{Arxiv}, 2022.

\bibitem{dualvc3}
Z.~Ning, S.~Wang, P.~Zhu, Z.~Wang, J.~Yao, L.~Xie, and M.~Bi, ``Dualvc 3: Leveraging language model generated pseudo context for end-to-end low latency streaming voice conversion,'' \emph{ArXiv}, 2024.

\bibitem{dualvc}
Z.~Ning, Y.~Jiang, P.~Zhu, J.~Yao, S.~Wang, L.~Xie, and M.~Bi, ``Dualvc: Dual-mode voice conversion using intra-model knowledge distillation and hybrid predictive coding,'' in \emph{International Speech Communication Association (Interspeech)}, 2023, pp. 2063--2067.

\bibitem{dualvc2}
Z.~Ning, Y.~Jiang, P.~Zhu, S.~Wang, J.~Yao, L.~Xie, and M.~Bi, ``Dualvc 2: Dynamic masked convolution for unified streaming and non-streaming voice conversion,'' in \emph{International Conference on Acoustics, Speech and Signal Processing (ICASSP)}, 2024, pp. 11\,106--11\,110.

\bibitem{fasts2s}
H.~Kameoka, K.~Tanaka, and T.~Kaneko, ``Fasts2s-vc: Streaming non-autoregressive sequence-to-sequence voice conversion,'' \emph{Arxiv}, 2021.

\bibitem{investgateStreaming}
T.~Hayashi, K.~Kobayashi, and T.~Toda, ``An investigation of streaming non-autoregressive sequence-to-sequence voice conversion,'' in \emph{International Conference on Acoustics, Speech and Signal Processing (ICASSP)}, 2022, pp. 6802--6806.

\bibitem{PPGSun2016PhoneticPF}
L.~Sun, K.~Li, H.~Wang, S.~Kang, and H.~Meng, ``Phonetic posteriorgrams for many-to-one voice conversion without parallel data training,'' in \emph{International Conference on Multimedia and Expo (ICME)}, 2016, pp. 1--6.

\bibitem{nansy}
H.-S. Choi, J.~Lee, W.~Kim, J.~Lee, H.~Heo, and K.~Lee, ``Neural analysis and synthesis: Reconstructing speech from self-supervised representations,'' in \emph{Neural Information Processing Systems(NeurIPS)}, 2021, pp. 16\,251--16\,265.

\bibitem{mediumvc}
Y.~Gu, Z.~Zhang, X.~Yi, and X.~Zhao, ``Mediumvc: Any-to-any voice conversion using synthetic specific-speaker speeches as intermedium features,'' \emph{Arxiv}, 2021.

\bibitem{wang2023multilevel}
Z.~Wang, L.~Xue, Q.~Kong, L.~Xie, Y.~Chen, Q.~Tian, and Y.~Wang, ``Multi-level temporal-channel speaker retrieval for zero-shot voice conversion,'' \emph{IEEE/ACM Transactions on Audio, Speech, and Language Processing}, vol.~32, pp. 2926--2937, 2024.

\bibitem{INchou2019oneshot}
J.~chieh Chou and H.-Y. Lee, ``One-shot voice conversion by separating speaker and content representations with instance normalization,'' in \emph{International Speech Communication Association (Interspeech)}, 2019, pp. 664--668.

\bibitem{VQMIVC}
D.~Wang, L.~Deng, Y.~T. Yeung, X.~Chen, X.~Liu, and H.~Meng, ``Vqmivc: Vector quantization and mutual information-based unsupervised speech representation disentanglement for one-shot voice conversion,'' in \emph{International Speech Communication Association (Interspeech)}, 2021, pp. 1344--1348.

\bibitem{contrastive}
J.~Ebbers, M.~Kuhlmann, T.~Cord-Landwehr, and R.~Haeb-Umbach, ``Contrastive predictive coding supported factorized variational autoencoder for unsupervised learning of disentangled speech representations,'' in \emph{International Conference on Acoustics, Speech and Signal Processing (ICASSP)}, 2021, pp. 3860--3864.

\bibitem{avqvc}
H.~Tang, X.~Zhang, J.~Wang, N.~Cheng, and J.~Xiao, ``Avqvc: One-shot voice conversion by vector quantization with applying contrastive learning,'' in \emph{Conference on Acoustics, Speech and Signal Processing (ICASSP)}, 2022, pp. 4613--4617.

\bibitem{liveosvc}
H.~Yang, L.~Deng, Y.~T. Yeung, N.~Zheng, and Y.~Xu, ``Streamable speech representation disentanglement and multi-level prosody modeling for live one-shot voice conversion,'' in \emph{International Speech Communication Association (Interspeech)}, 2022, pp. 2578--2582.

\bibitem{ALO-VC}
B.~Wang, D.~Ronssin, and M.~Cernak, ``Alo-vc: Any-to-any low-latency one-shot voice conversion,'' in \emph{International Speech Communication Association (Interspeech)}, 2023, pp. 2073--2077.

\bibitem{streamvoice}
Z.~Wang, Y.~Chen, X.~Wang, L.~Xie, and Y.~Wang, ``Streamvoice: Streamable context-aware language modeling for real-time zero-shot voice conversion,'' \emph{Arxiv}, 2024.

\bibitem{makeavoice}
R.~Huang, C.~Zhang, Y.~Wang, D.~Yang, L.~Liu, Z.~Ye, Z.~Jiang, C.~Weng, Z.~Zhao, and D.~Yu, ``Make-a-voice: Unified voice synthesis with discrete representation,'' \emph{Arxiv}, 2023.

\bibitem{uniaudio}
D.~Yang, J.~Tian, X.~Tan, R.~Huang, S.~Liu, X.~Chang, J.~Shi, S.~Zhao, J.~Bian, X.~Wu \emph{et~al.}, ``Uniaudio: An audio foundation model toward universal audio generation,'' \emph{Arxiv}, 2023.

\bibitem{PEFT}
Z.~Han, C.~Gao, J.~Liu, S.~Q. Zhang \emph{et~al.}, ``Parameter-efficient fine-tuning for large models: A comprehensive survey,'' \emph{Arxiv}, 2024.

\bibitem{tang2024salmonn}
C.~Tang, W.~Yu, G.~Sun, X.~Chen, T.~Tan, W.~Li, L.~Lu, Z.~MA, and C.~Zhang, ``{SALMONN}: Towards generic hearing abilities for large language models,'' in \emph{International Conference on Learning Representations (ICLR)}, 2024.

\bibitem{ma2024embarrassingly}
Z.~Ma, G.~Yang, Y.~Yang, Z.~Gao, J.~Wang, Z.~Du, F.~Yu, Q.~Chen, S.~Zheng, S.~Zhang \emph{et~al.}, ``An embarrassingly simple approach for llm with strong asr capacity,'' \emph{Arxiv}, 2024.

\bibitem{lora}
E.~J. Hu, yelong shen, P.~Wallis, Z.~Allen-Zhu, Y.~Li, S.~Wang, L.~Wang, and W.~Chen, ``Lo{RA}: Low-rank adaptation of large language models,'' in \emph{International Conference on Learning Representations (ICLR)}, 2022.

\bibitem{audiodec}
Y.-C. Wu, I.~D. Gebru, D.~Markovi{\'c}, and A.~Richard, ``Audiodec: An open-source streaming high-fidelity neural audio codec,'' in \emph{International Conference on Acoustics, Speech and Signal Processing (ICASSP)}, 2023, pp. 1--5.

\bibitem{transformer}
A.~Vaswani, N.~Shazeer, N.~Parmar, J.~Uszkoreit, L.~Jones, A.~N. Gomez, L.~Kaiser, and I.~Polosukhin, ``Attention is all you need,'' in \emph{International Conference on Neural Information Processing Systems(NIPS)}, 2017, p. 6000–6010.

\bibitem{vediollama2}
Z.~Cheng, S.~Leng, H.~Zhang, Y.~Xin, X.~Li, G.~Chen, Y.~Zhu, W.~Zhang, Z.~Luo, D.~Zhao, and L.~Bing, ``Videollama 2: Advancing spatial-temporal modeling and audio understanding in video-llms,'' \emph{Arxiv}, 2024.

\bibitem{repcodec}
Z.~Huang, C.~Meng, and T.~Ko, ``Repcodec: A speech representation codec for speech tokenization,'' \emph{Arxiv}, 2024.

\bibitem{intermedium}
C.~Wang, Y.~Wu, S.~Chen, S.~Liu, J.~Li, Y.~Qian, and Z.~Yang, ``Improving self-supervised learning for speech recognition with intermediate layer supervision,'' in \emph{International Conference on Acoustics, Speech and Signal Processing (ICASSP)}, 2022, pp. 7092--7096.

\bibitem{spearTTS}
E.~Kharitonov, D.~Vincent, Z.~Borsos, R.~Marinier, S.~Girgin, O.~Pietquin, M.~Sharifi, M.~Tagliasacchi, and N.~Zeghidour, ``{Speak, Read and Prompt}: High-fidelity text-to-speech with minimal supervision,'' \emph{ArXiv}, 2023.

\bibitem{wenetspeech4tts}
L.~Ma, D.~Guo, K.~Song, Y.~Jiang, S.~Wang, L.~Xue, W.~Xu, H.~Zhao, B.~Zhang, and L.~Xie, ``Wenetspeech4tts: A 12,800-hour mandarin tts corpus for large speech generation model benchmark,'' \emph{Arxiv}, 2024.

\bibitem{aishell3}
Y.~Shi, H.~Bu, X.~Xu, S.~Zhang, and M.~Li, ``{AISHELL-3}: A multi-speaker mandarin tts corpus,'' in \emph{International Speech Communication Association (Interspeech)}, 2021, pp. 2756--2760.

\bibitem{wenetspeech}
B.~Zhang, H.~Lv, P.~Guo, Q.~Shao, C.~Yang, L.~Xie, X.~Xu, H.~Bu, X.~Chen, C.~Zeng \emph{et~al.}, ``Wenetspeech: A 10000+ hours multi-domain mandarin corpus for speech recognition,'' in \emph{International Conference on Acoustics, Speech and Signal Processing (ICASSP)}, 2022, pp. 6182--6186.

\bibitem{didispeech}
T.~Guo, C.~Wen, D.~Jiang, N.~Luo, R.~Zhang, S.~Zhao, W.~Li, C.~Gong, W.~Zou, K.~Han, and X.~Li, ``Didispeech: A large scale mandarin speech corpus,'' in \emph{International Conference on Acoustics, Speech and Signal Processing (ICASSP)}, 2021, pp. 6968--6972.

\bibitem{emime}
M.~Wester, ``The {EMIME} bilingual database,'' The University of Edinburgh, Tech. Rep., 2010.

\bibitem{llama}
H.~Touvron, T.~Lavril, G.~Izacard, X.~Martinet, M.-A. Lachaux, T.~Lacroix, B.~Rozi{\`e}re, N.~Goyal, E.~Hambro, F.~Azhar \emph{et~al.}, ``Llama: Open and efficient foundation language models,'' \emph{Arxiv}, 2023.

\bibitem{wespeaker}
H.~Wang, C.~Liang, S.~Wang, Z.~Chen, B.~Zhang, X.~Xiang, Y.~Deng, and Y.~Qian, ``Wespeaker: A research and production oriented speaker embedding learning toolkit,'' in \emph{International Conference on Acoustics, Speech and Signal Processing (ICASSP)}, 2023, pp. 1--5.

\end{thebibliography}

\end{document}